# Automated identification of neural cells in the multi-photon images using deep-neural networks

Si-Baek Seong[1] and Hae-Jeong Park[1,2,3]

[1] BK21 PLUS Project for Medical Science, Yonsei University College of Medicine
[2] Department of Nuclear Medicine, Yonsei University College of Medicine
[3] Center for Systems and Translational Brain Sciences, Institute of Human Complexity and Systems Science, Yonsei University, Seoul, Republic of Korea
sbseong@yuhs.ac, parkhj@yonsei.ac.kr

**Abstract.**
The advancement of the neuroscientific imaging techniques has produced an unprecedented size of neural cell imaging data, which calls for automated processing. In particular, identification of cells from two-photon images demands segmentation of neural cells out of various materials and classification of the segmented cells according to their cell types. To respond the increasing demands in the neuroscience, the current study was conducted to provide an automated scheme for neural cell identification from combinations of calcium images (syn-GCaMP6s) and mCherry fluorescence images detected using two-photon imaging. To automatically segment neural cells, we used U-Net model, followed by classification of excitatory and inhibitory neurons and glia cells using a transfer learning technique. For transfer learning, we tested three public models of resnet18, resnet50 and inceptionv3, after replacing the fully connected layer with that for three classes. The best classification performance was found for the model with inceptionv3. The proposed application of deep learning technique is expected to provide a critical way to cell identification in the era of big neuroscience data.

## 1   Introduction

The advanced neuroscientific imaging techniques have produced an unprecedented amount of neural imaging data. For example, multi-photon calcium imaging (CaI) becomes an important tool to assess activity of neural population [1]. The temporal activity measured in CaI allows exploration of the functional tuning of each neuron with a relatively high temporal resolution. This is generally done after identifying neural types for example, excitatory and inhibitory and glia cells. Recent optogenetic techniques make it possible to differentiate inhibitory and excitatory neurons and glial cells [2] using two-photon imaging. A large number of neural cells measured using a high-resolution optical imaging calls for an automated method to identify neural cells. The identification of neural cells begins with segmentation of cells, followed by classification of the segmented cells according to their cell types.

In the segmentation and classification of images, deep learning techniques have shown a remarkable performance in various domains. It has been used for segmentation of various types of images and classification diverse objects in various applications. The application of deep learning to cell segmentation [3] and cell classification [4] have been introduced recently [5]. Among many deep learning architectures for segmentation, U-Net is one of widely used public network models [3, 5], conducting convolution and deconvolution in the form of U shape to train the upsampling process while reflecting the information obtained from the forward network [3].

The successful application area of the deep learning is the classification. There are many models that are trained using a large sized database of natural images, for example, Alexnet [6], GoogLeNet [7], ResNet [8] and Inception-v3 [9]. In spite of successful stories for those models in the natural image classification, the direct application of those models or architecture to cell classification is not generally efficient due to insufficient data to train a deep neural network. Instead of training a new model with a limited data set, it is possible to reuse pretrained models in the application of new problem area, which is called transfer learning . Previous studies have shown the excellent performance of the transfer learning in the specific applications, not restricted to the natural scene classification, with a relatively fewer data size. This transfer learning method has been used to identify multiple cells [10]. In the transfer learning, in most cases, we need to re-train only the fully connected layer before the final softmax layer with a data in the new application domain.

So far, several studies have been conducted to segment and to classify cells. Apthorpe proposes a CNN structure for detecting neurons in a representative image, and uses a method of creating a patch immediately by eliminating the redundancy of each neuron [11]. This approach trains the parameters of the network with the labeled representative image and does not include the sequential information. Kim applies a U-Net architecture to the image he created with the patch to produce an output that determines whether it is a cell or not [12].

To respond the increasing demands in the neuroscience, the current study was conducted to provide an automated pipeline for neural cell identification using a series of deep learning algorithms. For the identification of neural cell types in a two-phone imaging, we used automated segmentation using U-net and adopted a transfer learning technique for the classification of the segmented regions. For transfer learning, we reused ResNet18[13], ResNet50 [8] and Inception-v3[9]. We reused model weights lower than the fully connected network and re-trained the fully connected network with the two-phone images. To evaluate the performance, we compared the current segmentation with a previously done algorithm available from the public database. We also conducted 10-fold cross-validation for the classification performance. The contribution of the current study is as below.

## 2  Methods

### 2.1  Data description

For the neural cell identification, we used a dataset of mouse barrel cortex, available from a public database (https://crcns.org/data-sets/ssc/ssc-2) [2, 14]. The dataset contains six 6–8-week old emx1-Cre X LSL-H2B-mCherry mice, which had red nuclei in excitatory neurons and glial cells [15]. They were infected with AAV2/1 syn-GCaMP6s [16] to indicate the concentration of cytosolic calcium ions in neurons. The syn-GCaMP6s positive indicates that the cell is a neuron [16]. H2B-mCherry positive (brighter inside of a cell than the surrounding) indicates that the cell is an excitatory neuron or a glial cell [15]. Green (GCaMP; BG22) and red (mCherry; 675/70 emission filter, Chroma) fluorescence channels were acquired at the same time using a two-photon microscope images with a 16X 0.8 NA objective (Nikon). Three 600 by 600 μm (512 by 512 pixels) imaging planes attained with separation by 15 um in depth. Total 126 images from seven mice were used. mCherry and GCaMP images were combined to compose an RGB image, by assigning them to red and green channels.

To train a segmentation model, we used a probability map for each cell in the database, which was derived by a greedy template fitting algorithm applied to the average image and time information of the calcium signals [2]. For classification, three medical students labeled patches of 5000 cells in two-photon images to one of three types (excitatory, inhibitory neurons and glial cells). We used 2500 cell patches that all three people evaluated with the same label.

### 2.2  Image segmentation using U-Net architecture

For the segmentation, we added the two channel images (mCherry and GCaMP) after intensity normalization to make a grayscale image (Fig. 1. B), which was used as input to U-Net [3, 5] architecture. We used cell presence probability maps, which were derived by a greedy template fitting algorithm [2], as a target label. A total of 126 images were used for training. The models were trained for 10 epochs with 1000 iterations per epoch. As a target function, the probability of a cell predicted through a network is compared with the probability(p) of labels (p=1.0) in pixel units, and the accuracy is shown. The parameter is trained through a cross-entropy function. The U-Net architecture utilizes an overlap tile strategy to fine-tune deconvolution and uses elastic deformation for data augmentation (Fig. 2.).

### 2.3  Classification model: Residual Network and Inception-v3 network

The segmented image was thresholded with a threshold value of 0.7 and was underwent a binary object extraction process. In the binary object extraction algorithm, the clustering was done based on the connection level 8 (connected to all the surrounding pixels from a pixel). A patch of 101x101 units for each cluster from the centroid of the cluster is generated, followed by upsampling from 101x101 pixels to 299x299 pixels for the input of the transfer learning.

The data augmentation was performed for training with pixel angle conversion of (-30, 30) degrees and with scale conversion of (0.9, 1.1) to amplify the number of the data twice. Finally, we used 4500 patches for training.

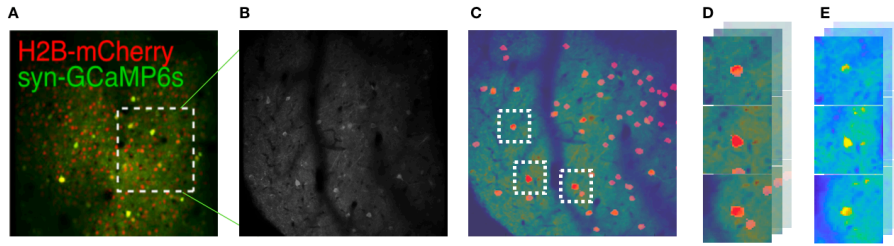

**Fig. 1.** Overview of getting patches from overlay labeled data and segmented images A) original image with both photon image. B) `gray scaled image`. C) Image overlaid with segment image filtered to threshold(>0.8) in original gray image. Make a patch with each cell over threshold. D) find the cell in segmented image and E) are original image patches to use classification model.

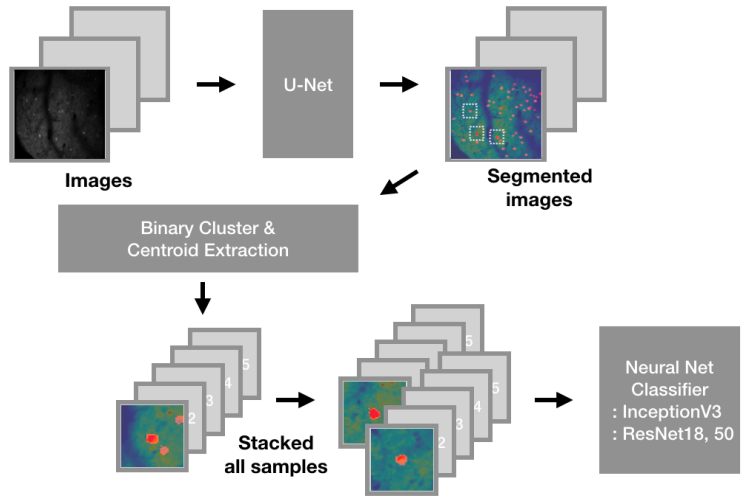

**Fig. 2.** Overview of whole process. Grayscale images are segmented through U-Net, and the clusters are searched for using the binary cluster method, and images containing cells are generated based on the center points. After the image is preprocessed, it passes through the classifier network to distinguish the cells.

Using the centroid coordinates from segmentation, red and green channel patches were obtained from mCherry and GCaMP. To make a blue channel, we assigned the average of normalized mCherry and GCaMP. Upon a pre-training model, we replace the fully connected layer to generate three outputs using softmax output functions.

**Multi-point re-train model evaluation.** For image classification, we used two base models of GoogLeNet and ResNet. When transfer learning is conducted, most studies re-train only the last fully connected (FC) layer. This is because the early layers are considered to capture local features, which the later layers classify into multiple classes for the application. However, in order to optimize the transfer learning, we compared the model performance by allowing parameter training from pointed layers (add or mixed layer) to last FC layer in the two base models. For training, 8 (ResNet) to 11(inceptionNet) epochs were generally needed to arrive at the saturation point.

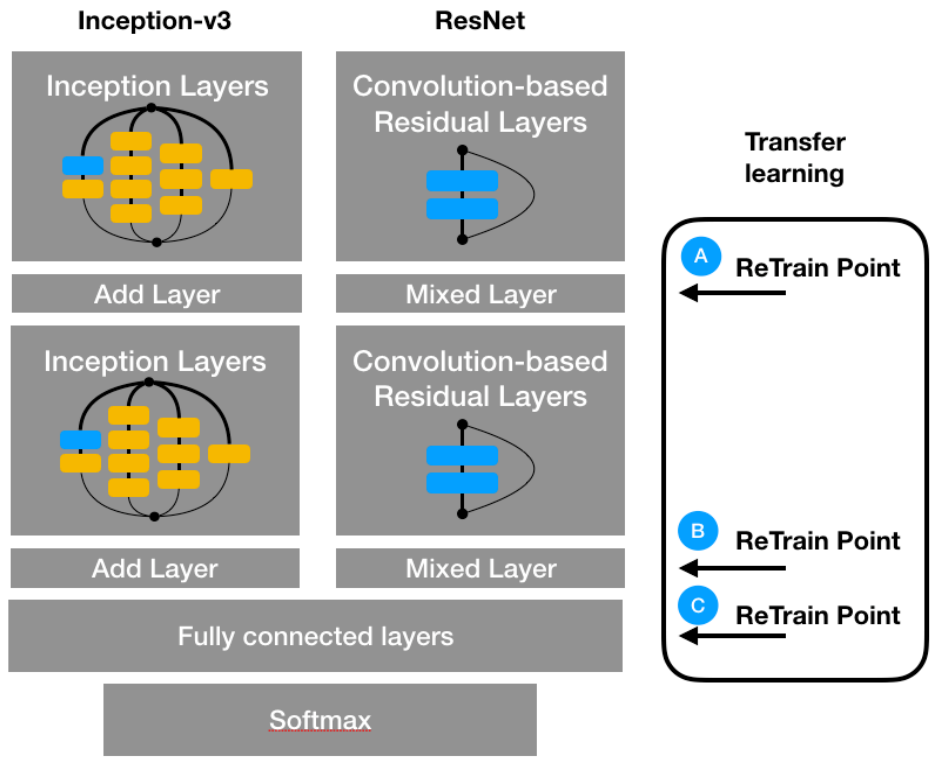

**Fig. 3.** A schematic diagram of multi-point re-train model. Re-training is performed from the arrowed area (mixed or Add layer) to the fully connected layer. When selecting A-model, network re-train from A to C. When selecting B-model, network re-train from B to C. In the figure, only each unit structure was included, but proceeded in descending order.

## 3   Results

The U-Net trained with barrel cortex representative 126 images (containing about 5000 cells) showed a segmentation accuracy of 93.2% when we considered greedy template fitting results as a ground-truth [Fig. 4A]. The classification results for various models were presented in Fig. 4B and 4C. For three different cell types, i.e., glia cells, inhibitory neurons, and excitatory neurons, the mean accuracy are presented in Table 1. A confusion matrix for classification was presented in Table 2. The best accuracy for six models of ResNet18 and 50 (Table 1. 1-6) were 82.93 ± 4.45, 85.93 ± 5.43, 84.89 ± 5.18, 90.79 ± 3.97, 92.49 ± 2.91, and 93.45 ± 3.37 respectively. In the model of Inception-v3 network, the accuracy were 79.58 ± 11.61, 87.64 ± 6.99, 93.12 ± 4.47, and 96.17 ± 3.75 according to four different models (Table 1. 7-10).

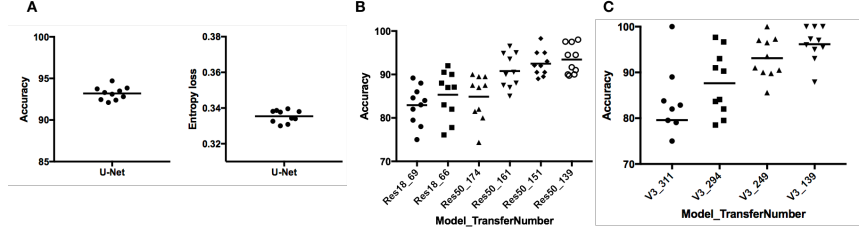

**Fig. 4.** Results of all networks. A) Segmentation result (mean accuracy: 93.2%). B) classification accuracy for Residual Network evaluated using a 10-fold cross validation. ResNet50 models has better performance than ResNet18 models. The number in the parenthesis for each model indicates the lowest layer (up to the fully connected layer) allowed to update weights using the new data. For example, in the case of ResNet18_69, the accuracy is the result of re-training the layers 70 and above using the new data. In this case, we did not train lower layers from layers 1 to 69 (Fig. 4. B.) The best results of ResNet were obtained from training ResNet50 from 139 layers. C) Classification accuracy for Inceptionv3 models. InceptionV3 with the 139 th layer as the lowest trainable layer model showed best performance in the whole models.

**Table 1.** Mean and standard deviation for saturated epoch accuracy of network models. Model architecture name with the lowest layer allowed to update weights during the training stage were presented. Best accuracy is shown in bold.

| No. | Methods (Transfer Layer) | Mean Accuracy (saturation) | Mean Accuracy (best epoch) | Standard Deviation | Saturation Epoch |
|---|---|---|---|---|---|
| 1 | U+ResNet18 (69) | 80.630 | 80.634 | 6.701 | 8 |
| 2 | U+ResNet18 (66) | **82.190** | **82.930** | 5.536 | 8 |
| 3 | U+ResNet50(174) | 81.370 | 84.886 | 4.635 | 8 |
| 4 | U+ResNet50(161) | 87.880 | 90.792 | 5.424 | 8 |
| 5 | U+ResNet50(151) | 86.860 | 92.493 | 5.583 | 8 |
| 6 | U+ResNet50(139) | **88.950** | **93.447** | 6.065 | 8 |
| 7 | U+Inception-v3 (311) | 77.370 | 79.580 | 11.61 | 11 |
| 8 | U+Inception-v3 (294) | 84.670 | 87.640 | 6.986 | 11 |
| 9 | U+Inception-v3 (249) | 87.200 | 93.120 | 4.473 | 11 |
| 10 | U+Inception-v3 (139) | **91.250** | **96.170** | 4.792 | 11 |

**Table 2.** Confusion matrix of sum of 10-fold validation set results at 11 epoch results of saturation model (Inception-v3(139)). Excitatory neurons and glia cells had more samples than inhibitory neurons.

|  | Excitatory | Glial cell | Inhibitory | Sens. / speci. |
|---|---|---|---|---|
| Excitatory | 159 | 4 | 13 | 90.34 |
| Glial cell | 1 | 185 | 3 | 97.88 |
| Inhibitory | 12 | 1 | 21 | 61.76 |
| Accuracy | - | - | - | **91.48** |

## 4    Conclusion

In this paper, we proposed a method for detecting neurons in calcium images without using large amounts of time series information. The novelty of our study was to identify the neurons rather than glia in the calcium image by ensemble of segmentation and classification algorithms, and it was able to find out whether it was a particularly excitatory neuron or inhibitory neuron. In addition, it is possible to solve the uniformity problem in the patch in the deep learning study by combining the segmentation and classification. If transfer learning is performed using ResNet and Inception network, performance can be improved by effective training when studying models based on mixed layer and add layer (Fig. 3). Additional training and classification models can be created using a small amount of memory. The above models can train each 10-fold model using GTX1080, except for inception-v3 (139). It is important to select and train the learning points without training all the layers.

Existing cell classification methods were trained and classified by making a patch consisting of only one cell [4, 17]. This was weak to noise if other cells came into the patch. In addition, there is a phenomenon in which some cytoplasm of the target cell is also cleaved to prevent other cell nuclei from penetrating into the patch [4]. Due to the nature of the calcium image, it takes a lot of time to preprocess one cell. Unlike the existing methodology, our model allows us to use existing data just after segmentation process. This proposed application of deep learning technique is expected to provide a critical way to cell identification in the era of big neuroscience data.

**Acknowledgments**

This research was supported by Brain Research Program and the Korea Research Fellowship Program through the National Research Foundation of Korea (NRF) funded by the Ministry of Science and ICT (NRF-2017M3C7A1049051).